# Robust estimation of inequality from binned incomes

Paul T. von Hippel (University of Texas, Austin)
Samuel V. Scarpino (Santa Fe Institute)
Igor Holas (University of Texas, Austin)

*Correspondence*: Paul T. von Hippel, LBJ School of Public Affairs, University of Texas, 2315 Red River, Box Y, Austin, TX 78712, paulvonhippel.utaustin@gmail.com.

*Contributors*: Von Hippel conceived the study, led the writing, and developed and ran the Stata code. Scarpino developed and ran the R code. Holas summarized the results.

*Acknowledgments*: von Hippel and Holas were supported in part by a New Scholar Grant from the Stanford Center on Poverty and Inequality and by a grant from the Policy Research Institute at the University of Texas. Scarpino was supported by a National Science Foundation Graduate Research Fellowship, by the Santa Fe Institute, and by the Omidyar Group.

*Bios*: Paul T. von Hippel's research interests include family and school influences on academic achievement. With Holas and Scarpino, he has applied the methods here to another paper titled "Secession of the Successful or Millionaires Next Door? Trends in family income inequality, within and between school districts, 1970-2009."

Samuel V. Scarpino is an Omidyar Fellow at the Santa Fe Institute and an incoming Assistant Professor in the Department of Mathematics and Statistics and the Complex Systems Center at the University of Vermont. His research primarily focuses on the evolutionary and population dynamics of infectious diseases. Sam earned a Ph.D. in Integrative Biology from the University of Texas at Austin.

Igor Holas is a PhD candidate in Human Development and Family Science at the University of Texas, and co-founder and CEO of Mentegram, which helps mental health professionals to provide remote monitoring and remote interventions.

# Robust estimation of inequality from binned incomes

*Abstract*

Researchers must often estimate income inequality using data that give only the number of cases (e.g., families or households) whose incomes fall in “bins” such as $0-9,999, $10,000-14,999,…, $200,000+. We find that popular methods for estimating inequality from binned incomes are not robust in small samples, where popular methods can produce infinite, undefined, or arbitrarily large estimates. To solve these and other problems, we develop two improved estimators: the robust Pareto midpoint estimator (RPME) and the multimodel generalized beta estimator (MGBE). In a broad evaluation using US national, state, and county data from 1970 to 2009, we find that both estimators produce very good estimates of the mean and Gini, but less accurate estimates of the Theil and mean log deviation. Neither estimator is uniformly more accurate, but the RPME is much faster, which may be a consideration when many estimates must be obtained from many datasets. We have made the methods available as the *rpme* and *mgbe* commands for Stata and the *binequality* package for R.

# 1. INTRODUCTION

Social scientists often wish to estimate income inequality, not just for nations and states, but within small areas such as neighborhoods, school districts, and counties. There are numerous examples in sociology and economics:

- Jargowsky (1996) estimated 20-year trends in the segregation of metropolitan neighborhoods by income. To estimate his neighborhood segregation index, he had to estimate the mean and variance of income within each neighborhood.
- Corcoran and Evans (2010) asked whether school districts with greater income inequality were more likely to vote for redistribution in the form of high local school taxes. To estimate inequality within districts, the authors had to estimate the mean and median income within each school district.
- Galbraith and Hale (2004, 2009) estimated 40-year trends in the component of the Theil inequality statistic that lies between US counties. It would be informative to compare this to trends in the Theil component that lies within counties.

All these studies of small-area inequality encountered the same practical challenge. The challenge is that, within small areas, researchers rarely have data on the incomes of individual *cases* (e.g., households or families). Instead, researchers must estimate the distribution of incomes from income *bins* (also known as brackets, groups, or intervals). As an example of binned income data, Table 1 summarizes the distribution of household incomes within the richest county and the poorest county in the United States: Nantucket County, Massachusetts, and Maricao County, Puerto Rico. For each county, Table 1 estimates how many households have incomes in each of 16 bins. The top bin $[\$200{,}000, \infty)$ is one-sided because it has no upper bound. Technically, the bottom bin is also one-sided because it has no lower bound, but it is customary and innocuous to treat the bottom bin as though its lower bound is zero—i.e., [$0, $10,000).[1]

It is clear that Nantucket is much richer than Maricao. Yet it is challenging to be more specific. How much inequality lies between the two counties—for example, how many times larger is the mean or median household income in Nantucket than in Maricao? And how do the two counties differ with respect to a measure of within-county inequality such as the Gini coefficient, the Theil index, the coefficient of variation (CV), or the mean log deviation (MLD)? Questions like these are fundamental to research on small-area inequality, and their answers must often be estimated from binned data.

A variety of methods have been proposed for estimation with binned incomes. In fact, much more effort has been spent on developing estimators than on evaluating their performance, especially in small samples. Most methodological articles simply describe a binned-data estimator and then demonstrate its use in a handful of datasets, typically with large samples (for exceptions, see Bandourian, McDonald, & Turley, 2002; Evans, Hout, & Mayer, 2004; Minoiu & Reddy, 2009, 2012; von Hippel, Holas, & Scarpino, 2012). The limited nature of past evaluations was in part due to the limited availability of published inequality statistics, especially for small areas. Until recently, no inequality statistics were available below the level of US states, and even at the state level no statistic was available except for the Gini.

It is now possible to more thoroughly evaluate binned-data estimators in small samples, since the US Census Bureau has published the household Gini of all 3,221 US counties and the household Gini, Theil, and mean log deviation (MLD) of every US state (Bee, 2012; Hisnanick & Rogers, 2006). In this article, we use these and other statistics to carry out the largest evaluation of binned-data estimators that has been conducted to date.

Our evaluation focuses on two of the most popular estimators. One is the simple Pareto midpoint estimator (PME) (Henson, 1967). The PME assigns cases to the midpoint of their bins, except for the top bin, where there is no midpoint and the PME assigns cases to the arithmetic mean of a Pareto distribution. The other method involves fitting distributions from the generalized beta (GB) family, which includes the Pareto, log normal, log logistic, Dagum, and Singh-Maddala distributions, as well as different forms of the beta and gamma distributions (McDonald & Xu, 1995).

Our evaluation uncovers problems that have escaped notice in more limited evaluations. In particular, we find that neither estimator, as commonly implemented, is robust in small samples. In small samples, the PME sometimes yields estimates that are arbitrarily large or even infinite. Some GB distributions can also yield infinite or undefined estimates, at least on occasion, and some can fail to yield an estimate at all because of nonidentification or nonconvergence. In addition, past literature gives incomplete guidance regarding which distribution from the GB family should be fit to a particular dataset.

We modify both estimators to make them more robust in small samples. To improve the PME, we develop a *robust Pareto midpoint estimator* (RPME), which is just like the PME except that the mean of the top bin is guaranteed to be defined and not arbitrarily large. We offer several flavors of the RPME, which achieve robustness by constraining the Pareto shape parameter and/or by replacing the arithmetic mean with a robust alternative such as the harmonic mean. These changes make little difference in large samples but dramatically improve the estimates in small samples.

To improve the utility of the GB family, we develop a *multimodel GB estimator* (MGBE). The MGBE fits 10 distributions from the GB family, discards any fitted distributions with undefined estimates, and selects or averages among the remaining distributions according to some criterion of fit. By default our criterion is the Akaike Information Criterion (AIC), but we get very similar results using the Bayes Information Criterion (BIC).

Between the RPME and the MGBE, neither is uniformly more accurate than the other. Theoretically, the RPME is a nearly nonparametric estimator that, with enough bins, can fit the nooks and crannies of any income distribution, regardless of its shape. RPME estimates can be very accurate if there are many bins, but much less accurate if there are few. The MGBE, by contrast, is a parametric estimator which assumes that incomes follow one of the distributions in the GB family. The GB family is quite flexible, but all of its member distributions are unimodal and positively skewed. If the true income distribution approximates one of the GB distributions, then the resulting estimates can be very accurate, even if the bins are coarse. However, if the true income distribution is a poor fit to the GB family—for example, if the true income distribution is bimodal—then estimates from the GB family will be biased (Minoiu & Reddy, 2009).

Empirically, the two estimators are about equally accurate in typical data with 8 to 20 bins. In artificial data with 4 bins the MGBE has a clear advantage, though both estimators are inaccurate. Both estimators are very good at estimating the mean, fairly good at estimating the Gini, and potentially poor at estimating the Theil and MLD. The MGBE is better than the RPME in estimating the median, though neither estimator is bad. The MGBE seems to be slightly more accurate in data from 1970-1990, while the RPME is slightly more accurate in data from 2000 and later. This could mean that the income distribution has changed in ways that worsen its fit to distributions in the GB family.

A further consideration is that the RPME is about 1,000 times faster than the MGBE. The difference in speed doesn't matter if you are fitting a single binned dataset, but if you are fitting thousands of binned datasets—for example, from every county or school district in the US—using the RPME can save hours.

Since either estimator can be better under different circumstances, when practical it would be wise for applied researchers to try both approaches and check the results against known reference statistics. To encourage adoption, we have made the methods available as the *rpme* command and *mgbe* commands for Stata, which can be installed using Stata's *ssc* command (Duan & von Hippel, 2016; von Hippel & Powers, 2014). Likewise, we have made both methods available in the *binequality* package for R, which can be downloaded from the Comprehensive R Archive Network at *http://cran.r-project.org/web/packages/binequality/index.html* (Scarpino, von Hippel, & Holas, 2014).

In the rest of this paper, we define the RPME and MGBE estimators in more detail, then present the data and results of our evaluation. We then draw conclusions from our evaluation and make final recommendations.

# 2. METHODS

In this section, we motivate and define the RPME and MGBE and related estimators.

## 2.1 Notation and terminology

Throughout this section we use the following notation and terminology. There are *n cases* (e.g., families or households) with different values of a continuous variable *X* (e.g., income or wealth). The distribution of *X* is summarized by a frequency table like Table 1 which gives the number of cases $n_b$ whose *X* values lie in each of several bins $[l_b, u_b)$, *b*=1,2,…. where $l_b$ and $u_b$ are the lower and upper bound of bin *b*.

In small samples or very poor populations it is not unusual for some of the bins to be *empty*, with $n_b = 0$, as the upper bins are in Maricao (Table 1). The empty bins contribute nothing to the estimates and can cause trouble for some calculations, so we will find it helpful to restrict the calculations to the *populated* bins with $n_b > 0$. The number of populated bins is called *B*. The total number of bins, both populated and unpopulated, will be called $B_{all}$.

## 2.2 The midpoint estimators

The most straightforward approach to binned data involves assigning cases to their bin midpoint. In this section we start with the most basic type of midpoint estimator, then review how the Pareto distribution can be used to estimate the mean of the top bin, and finally discuss how to make estimating from the top bin more robust.

### 2.2.1 The basic midpoint estimator (ME)

Within each bin, the basic midpoint estimator (ME) assigns the $n_b$ cases to the bin midpoint $m_b = (l_b + u_b)/2$ and then calculates statistics using the midpoints $m_b$. For example, the mean $\mu$ and variance $\sigma^2$ of *X* are estimated as

$$\begin{aligned}\hat{\mu}_{ME} &= \frac{1}{n}\sum_{b=1}^{B} n_b m_b \\ \hat{\sigma}^2_{ME} &= \frac{1}{n-1}\sum_{b=1}^{B} n_b (m_b - \hat{\mu}_{ME})^2\end{aligned} \qquad (1).$$

Naturally the ME is only defined if every populated bin has a defined upper and lower bound. In Table 1, for example, the ME is defined for Maricao but not for Nantucket.

Despite its simplicity, the ME has a very nice statistical property that we call *bin-consistency*. Bin-consistency means that, as the bins get narrower and more numerous, the ME estimate converges to its estimand. The ME is bin-consistent because it is nonparametric and makes no assumptions about the shape of the underlying distribution.

The properties of the ME get even nicer if certain assumptions are met. One assumption is that the distribution of *X* is "smooth" near the tails, which is a reasonable assumption but not one that can be verified from binned data. Another assumption is that the bin widths $w_b = u_b - l_b$ are roughly equal and not too much larger than the standard deviation $\sigma$ (Heitjan 1989 suggests $w_b < 1.6\sigma$). This assumption holds for some data but not for others. Consider Table 1. In Nantucket the populated bins are far from equal; they range in width from $5,000 to $50,000, but in Maricao the widest bins are empty and the populated bins are roughly in equal in width. So in Maricao the assumptions about bin width approximately hold, but in Nantucket they do not.

If the assumptions hold, $\hat{\mu}_{ME}$ will be nearly unbiased and $\hat{\sigma}^2_{ME}$ will have a bias that is correctable and typically small. The relative bias of $\hat{\sigma}^2_{ME}$ is approximately $(w_b/\sigma)^2/12$ (Heitjan, 1989). For example, in Maricao (Table 1) the populated bins are approximately $w_b \approx \sigma/2$ wide[2], so the relative bias of $\hat{\sigma}^2_{ME}$ is about 2%. That is, the estimate $\hat{\sigma}^2_{ME}$ is expected to be just 2% larger than the estimand $\sigma^2$.

In addition to having little bias, under the stated assumptions the estimates $\hat{\mu}_{ME}$ and $\hat{\sigma}^2_{ME}$ will be almost fully efficient. That is, they will be almost as efficient as the maximum likelihood estimates that would be obtained by fitting the binned data to the true distribution of *X*. This is a very nice property. Since we rarely know the true distribution of $X$, it is reassuring to know that knowing it would not greatly improve on the ME estimates.

The literature on the ME focuses on the estimands $\mu$ and $\sigma$, but if $\mu$ and $\sigma$ are well estimated then the inequality statistic $CV = \sigma/\mu$ will be well estimated, too. The Gini should also be well estimated since Gini = ½ $MD/\mu$ where MD is the mean difference which is closely related to $\sigma$ (Hosking, 1990). It is not clear how good the ME estimates will be for other inequality statistics such as the Theil and MLD.

### 2.2.2 *The Pareto midpoint estimator (PME)*

If the top bin is populated and has no upper bound, as in Nantucket (Table 1), then its midpoint is undefined, and some statistic other than the midpoint must be plugged in for the top bin. Intuitively the statistic should be some multiple of the top bin's lower bound $l_B$, and the multiple should be larger if there is evidence that the tail is longer. To estimate the length and shape of the tail, we must slightly compromise the nonparametric nature of the ME and assume that the incomes in the top two bins follow some parametric distribution.

An arbitrary but convenient and popular approach is to assume that the top two bins follow a Pareto distribution[3] with shape parameter $\alpha > 0$. Then the mean $\mu_B$ of the top bin is a simple function of $\alpha$

$$\mu_B = \begin{cases} l_B \dfrac{\alpha}{\alpha - 1} & if\ \alpha < 1 \\ \infty & if \alpha \geq 1 \end{cases} \qquad (2),$$

and an estimate of $\mu_B$ can be used in place of the midpoint. An estimate of $\mu_B$ can be obtained by plugging an estimate of $\alpha$ into the formula for $\mu_B$. The most popular estimator for $\alpha$ is

$$\hat{\alpha} = \frac{\ln((n_{B-1} + n_B)/n_B)}{\ln(l_B/l_{B-1})} = \frac{\ln(n_{B-1} + n_B) - \ln(n_B)}{\ln(l_B) - \ln(l_{B-1})} \qquad (3),$$

(Henson, 1967; Quandt, 1966). If the top two bins really follow a Pareto distribution, then $\hat{\alpha}$ is the maximum likelihood estimator. Past attempts to improve on the maximum likelihood estimator $\hat{\alpha}$ have not been successful (West, 1986; West, Kratzke, & Butani, 1992).[4]

### 2.2.3 *The robust Pareto midpoint estimator (RPME)*

Our results will show that the PME can perform very well in large samples, but it is not robust or even usable in some small samples. The problem is the sensitivity of $\mu_B$ to the value of $\alpha$. As equation (2) shows, if $\alpha \leq 1$ then the mean is infinite, and as $\alpha$ approaches 1 from above the mean grows arbitrarily large very quickly. In large samples, the estimate $\hat{\alpha}$ rarely gets into a problematic area, but in small samples, where $\hat{\alpha}$ has more error, $\hat{\alpha}$ can easily be close to or less than 1. For example, $\hat{\alpha}$ gets close to 0 as $n_{B-1}$ gets small compared to $n_B$.

Instead of the mean $\mu_B$ of the top bin, some authors have suggested using the median $M_B$ (Parker & Fenwick, 1983), and one could also use the geometric mean $g_B$ or the harmonic mean $h_B$. Under a Pareto distribution, these statistics are all simple functions of $\alpha$:

$$\begin{aligned} M_B &= l_B 2^{1/\alpha} \\ g_B &= l_B e^{1/\alpha} \\ h_B &= l_B(1 + 1/\alpha) \end{aligned} \tag{4}$$

(Johnson, Kotz, & Balakrishnan, 1994).

As shown in Figure 1, $M_B, g_B,$ and $h_B$ are far less sensitive to $\alpha$ than $\mu_B$ is. In particular, $M_B, g_B,$ and $h_B$ are defined for all (positive) values of $\alpha$, and $M_B, g_B,$ and $h_B$ do not get arbitrarily large until $\alpha$ gets close to 0. $h_B$ is the least sensitive to $\alpha$, and that makes $h_B$ the *a priori* best candidate for robust estimation. $h_B$ is also the smallest statistic when $\alpha < 1$, and that could introduce some negative bias, but the bias will be compensated by some reduction in variance—a tradeoff that we will evaluate empirically. For larger values of $\alpha$, the differences between the statistics are relatively small so the choice of statistic will matter less.

Since all the statistics can get arbitrarily large as $\alpha$ gets small, further robustness can be added by setting a lower bound $\alpha_{min}$ on the estimate of $\alpha$. That is, we can replace the estimate $\hat{\alpha}$ with

$$\tilde{\alpha} = \max(\alpha_{min}, \hat{\alpha}) \qquad (5),$$

and plug $\tilde{\alpha}$ into the formulas for $\mu_B, M_B, g_B,$ and $h_B$. If we are estimating the arithmetic mean, then $\alpha_{min}$ must be greater than 1, but if we are estimating one of the other statistics, then $\alpha_{min}$ need only be positive.

The best value for $\alpha_{min}$ must be estimated empirically, and that is a weakness to this approach. However, the $h_B$ statistic will be relatively insensitive to the value of $\alpha_{min}$ and will tolerate an $\alpha_{min}$ that is close to 0. That is another point in favor of using $h_B$ instead of one of the other statistics.

Our *rpme* command for Stata implements the RPME with a default of $h_B$ and $\alpha_{min} = 1$. There are options to change the value of $\alpha_{min}$ and an option to choose among $\mu_B, M_B, g_B,$ and $h_B$.

## 2.3 Continuous densities

A discomfiting feature of the RPME is that it models income as a discrete variable. We might hope to model income better by fitting a continuous density. Many continuous densities have been used to model income, although relatively few have been implemented in software that can fit binned incomes.

### 2.3.1 The generalized beta (GB) family

Some of the most popular densities for modeling income are members of the generalized beta (GB) family (McDonald, 1984; McDonald & Xu, 1995). Part of the GB family tree is drawn in Figure 1.[5] The tree starts with the flexible GB2 density, which has four positive parameters $\mu, \sigma, \nu, \tau$ (Stasinopoulos, Rigby, & Akantzilioutou, 2008)[6], and branches out into 3- and 2-parameter distributions including the log normal, log logistic, Pareto (type 2), beta (type 2), Dagum, Singh-Maddala, gamma and generalized gamma distributions. All of these distributions are either *nested* in the GB2 (as special cases that fix some parameters to a value of 1) or *embedded* in the GB2 (as limiting cases that are approached as some parameter goes to 0 or ∞).

Continuous densities can be fit to binned data by maximum likelihood (McDonald & Ransom, 2008). A binned-data maximum likelihood estimation of the GB family densities is available in R's *gamlss* package (Stasinopoulos et al., 2008), which we have used as the basis for our own *binequality* package (Scarpino et al., 2014). Formulas have been derived to transform maximum likelihood estimates of the GB parameters into estimates of the mean, variance, Gini, and Theil (McDonald & Ransom, 2008), but some of these formulas require functions that are difficult to implement, and there are no formulas for some statistics, including the MLD. A relatively easy and highly accurate alternative is to approximate the statistics using numerical methods (McDonald & Ransom, 2008). Our software implements a numerical recipe that draws 1,000 evenly spaced quantiles—i.e., the $0.05^{th}$, $0.15^{th}$, …, and $99.95^{th}$ percentiles—from the fitted distribution, and plugs them into sample formulas for statistical including the mean, median, variance, CV, Gini, Theil, and MLD.

### 2.3.2 Practical problems in fitting the GB family

A liability of GB family distributions is that they occasionally produce an undefined estimate of the mean or variance.[7] The mean of a GB family distribution is undefined whenever $-\hat{\nu} < 1/\hat{\sigma} < \hat{\tau}$ and the variance is undefined whenever $-\hat{\nu} < 2/\hat{\sigma} < \hat{\tau}$ (McDonald & Xu, 1995; Stasinopoulos et al., 2008). If the mean is undefined, all inequality statistics will be undefined as well, since all inequality statistics are functions of the mean. If the mean is defined but the variance is undefined, some inequality statistics will be undefined (such as the CV) and other inequality statistics may be very imprecise. Undefined estimates are more likely to occur in small samples, where parameter estimates are more variable and are more likely to wander into parts of the parameter space where the mean or variance is undefined.

Another issue that arises in fitting the GB family is *model uncertainty*. *A priori* it is not clear which distribution from the GB family will provide the best fit to the data. Past research has approached the problem of model uncertainty in several ways. One approach is to fit only the most general distribution, which is the 4-parameter GB2 (e.g., Chotikapanich, Rao, & Tang, 2007). This can produce good results in large samples, but it has several problems in small ones. First, the fitted GB2 distribution can have undefined moments, especially in small samples. Second, in small samples the 4 parameters of the GB2 distribution may be estimated imprecisely, so that the resulting estimates are worse than if a simpler 2- or 3-parameter distribution had been fit. Third, even in large samples, the GB2 distribution can only fit the distributions that it nests (the Dagum, Singh-Maddala, Beta 2, Pareto 2, and log logistic); it cannot fit perfectly the distributions that it embeds (the Weibull, log normal, gamma, and generalized gamma) because fitting those distributions would require a GB2 parameter to go to ∞ or 0.

Another approach to model uncertainty is to test the null hypothesis that the fitted density is in fact the true distribution of income. Bandourian et al. (2002) do this using a Pearson $X^2$ statistic; we do it using a likelihood ratio statistic (which is asymptotically equivalent to $X^2$):

$$G^2 = -2\ln(LR) = -2\left(\ell - \sum_{b=1}^{B} n_{\mathrm{b}}\left(\ln\left(\frac{n_b}{n}\right)\right)\right) \quad (6),$$

Here $\ell$ is the log likelihood of the fitted distribution. Under the null hypothesis that the fitted distribution is the true distribution, $G^2$ (or $X^2$) will follow an asymptotic chi-square distribution with

$$df = \min(B, B_{all} - 1) - k \quad (7),$$

degrees of freedom, where $k$ is the number of parameters, $B$ is the number of populated bins, and $B_{all}$ is all the bins including the empty ones (cf. Stirling, 1986). If all the bins are populated, this expression for degrees of freedom simplifies to $df=B-k-1$.[8]

The limitation of the null hypothesis test is that it rarely highlights a single distribution as fitting the data. In small samples, the power of the test is low, and there may be several distributions that it fails to reject. In large samples, the power of the test is high, and it may reject every model that is fit. This in fact what happened when Bandourian et al. (2002) fit GB family distributions to 23 countries in various years from 1967 to 1997. In every country and every year, they rejected every distribution in the GB family by a $X^2$ test.[9] In some countries, these rejections might have meant that the distributions fit poorly, but in other countries, it might have been that several distributions provided imperfect but very good approximations and might have produced good estimates of statistics like the Gini. A null hypothesis test cannot distinguish between these situations.

Another approach to model uncertainty is to decide between two models using a classical likelihood ratio (LR) test (McDonald, 1984). However, classical LR tests only apply to nested models (White, 1982) like the GB2 and Dagum; they cannot choose between embedded models like the GB2 and generalized gamma, or between models with the same number of parameters, such as the Dagum and Singh-Maddala. In addition, classical LR tests assume that one of the distributions being compared is in fact the true distribution of income (White, 1982). This is not a plausible assumption; as mentioned earlier, Bandourian et al. (2002) rejected the hypothesis that any distribution in the GB family was the true distribution of income in any of 23 countries.

### *2.3.3 The multimodel generalized beta estimator (MGBE)*

We now propose a multimodel generalized beta estimator (MGBE) which has several advantages over past approaches to the GB family. The MGBE fits every GB family distribution in Figure 2, discards any fitted distributions with an undefined variance (i.e., with $-\hat{\nu} < 2/\hat{\sigma} < \hat{\tau}$), and then uses *multimodel inference* to derive estimates from the remaining distributions.

Two criteria are possible, the Akaike Information Criterion (AIC) and the Bayes Information Criterion (BIC). We can use either AIC or BIC to *select* a single best model or to construct weights that we use to *average* estimates across models. Theoretically, model averaging should produce better estimates than model selection (Burnham & Anderson, 2004). Whether it is better to use AIC or BIC is a subtler issue that depends on how well the different GB family distributions actually fit the data (Burnham & Anderson, 2004), which is something that we can only assess empirically.

# 3. DATA AND RESULTS

In this section, we evaluate the RPME and MGBE in data on family and household incomes. We tune the estimators to data from US counties, then validate the tuned estimators in data from US states and from the US as a whole.

To evaluate the quality of an estimator, we would like to compare its estimates to the true value of the estimand in the population. For example, we would like to compare the mean and Gini estimated from binned incomes to the true Gini in the population. Unfortunately, the population Gini is unknown because the Census Bureau does not record income for every US family or household. So instead of population statistics, we use estimates calculated by the Census Bureau from unbinned incomes. Although these estimates are not population values, they are more accurate than binned-data estimates, and can therefore serve as a kind of gold standard. For brevity's sake we refer to them as "true" values or estimands in the evaluation.

We evaluate the accuracy of the estimates as follows. Let $\hat{\theta}_j$ and $\theta_j$ be the estimate and estimand for state or county $j$. Then $e_j = 100(\hat{\theta}_j - \theta_j)/\theta_j$ is the percent relative error of the estimate—i.e., the error expressed as a percent of the estimand. Across all the counties or states in the US, the *percent relative bias* is the mean of $e_j$, the *percent root mean squared error* (RMSE) is the square root of the mean of $e_j^2$, and the *reliability* of the estimate is the squared correlation between $\theta_j$ and $\hat{\theta}_j$. We also plot $\hat{\theta}_j$ against $\theta_j$ to visualize the accuracy of the estimates and see how the accuracy changes at different values of the estimand.

## 3.1 Counties in 2006-10

Our evaluation begins with household income data from each of the 3,221 counties in the US and Puerto Rico. We estimate the mean, median, and Gini from the county bins, and compare these estimates to the "true" values of the county mean, median, and Gini published by the Census (Bee, 2012; US Census Bureau, n.d.). The Census has never published county Ginis before.

The county data come from responses to the American Community Survey (ACS). To increase the number of households sampled per county, the ACS pooled data across the 5-year period from 2006 to 2010 to accumulate a 1-in-8 sample of US households with incomes adjusted to 2010 dollars. For each county, the Census summarized county incomes using 16 bins.

The binned data for two counties—Nantucket and Maricao—were given in Table 1. The counts in Table 1 are not sample counts but estimated population counts. Since the sampling fraction was 1 in 8, we approximated the sample counts by dividing the estimated population counts by 8. This makes no difference to the RPME but does slightly affect the MGBE since the AIC, BIC, and $G^2$ statistics are functions of sample size. The ACS is a complex sample survey, but unfortunately the binned data provide no information about the sample design, so we analyze the data as if they came from a simple random sample. This means that $p$ values etc. are only approximate (Rao & Scott, 1981).

County data presents a challenging estimation task because some counties have small populations, small sample sizes, empty bins, and lumpy income distributions. Maricao is a fairly small county, but some counties are much smaller; Loving County, Texas, is the smallest with only 31 households in 2010. County data will highlight the properties of our estimators with low bin counts, empty bins, and idiosyncratic income distributions.

### *3.1.1 RPME estimates*

We begin by tuning the RPME estimator. Figure 3 shows the RMSE and bias for estimates of the Gini and mean produced by different flavors of the RPME. There are flavors that use the arithmetic mean, the harmonic mean, the geometric mean, and the median; and for each flavor, $\alpha_{min}$ can take values from 0 to 4 or beyond. Note that the choice $\alpha_{min} = 0$ means that the estimate of $\alpha$ is not constrained at all.

Although nearly all past research used the arithmetic RPME with $\alpha_{min} = 0$—i.e., with no constraint on $\alpha_{min}$—that clearly and consistently produces the most biased estimates. It has an infinite RMSE when $\alpha_{min} < 1$ and a very large RMSE when $\alpha_{min}$ is greater than but close to 1. The arithmetic RPME can achieve a decent RMSE when $\alpha_{min} > 2$ but overall it is distressingly sensitive to the value chosen for $\alpha_{min}$.

The harmonic RPME is the most robust choice. It is the least sensitive to the specific value chosen for $\alpha_{min}$ and yields the smallest RMSE when $\alpha_{min} < 1$. It also yields a negative bias, but the bias is small when $\alpha_{min}$ is small, and the bias-variance tradeoff is favorable. The median and geometric RPME are also more robust than the arithmetic RPME, but not as robust as the harmonic RPME.

When the harmonic RPME is used, the most accurate estimates are obtained around $\alpha_{min} = 1$, although the precise value of $\alpha_{min}$ doesn't matter much. In these data, even $\alpha_{min} = 0$ looks like a plausible choice with the harmonic RPME, but in general setting $\alpha_{min} = 0$ is a bad idea since it could permit the estimates to get arbitrarily large.

$\alpha_{min} = 1$ is also a reasonable choice with the geometric RPME or the median RPME, but $\alpha_{min} = 2$ is a better choice with the arithmetic RPME. In addition to being empirically supported (at least in these data), all these choices have a common and intuitive interpretation.

They imply that the statistic used for the top bin cannot be more than twice that bin's lower bound.

Figure 4 illustrates the robustness that is achieved by using the harmonic RPME instead of the arithmetic RPME. The figure is a graphics grid that contains scatterplots of RPME-estimated county means against true county means. In the upper left, we see the RPME estimates obtained using the arithmetic RPME with $\alpha_{min} = 0$. Many of the estimates are much too large, and some are infinite. Increasing $\alpha_{min}$ to 1.1 gets rid of the infinite estimates, but still leaves many estimates that are much too large. Increasing $\alpha_{min}$ to 2 gives good results, but it is troubling that the results are so sensitive to the choice of $\alpha_{min}$.

By contrast, the lower row of Figure 4 shows the estimates produced by the harmonic RPME. The results are far more robust and much less sensitive to $\alpha_{min}$. The best results are achieved at $\alpha_{min} = 1.1$ but the results at $\alpha_{min} = 1$ or $\alpha_{min} = 2$ are only a little different.

Table 2 summarizes the accuracy of the arithmetic RPME with $\alpha_{min} = 2$ and the geometric, median, and harmonic RPMEs with $\alpha_{min} = 1$. In estimating the Gini, the arithmetic RPME is the most accurate, with the smallest bias, the smallest RMSE, and the highest reliability. However, the performance of the arithmetic RPME is highly dependent on the choice of $\alpha_{min}$, as we have seen. The harmonic RPME with $\alpha_{min} = 1$ provides estimates that are almost as good and less sensitive to the choice of $\alpha_{min}$. The geometric and median RPMEs are a little less accurate than the harmonic RPME, and more sensitive to $\alpha_{min}$. In estimating the mean, all 4 flavors of RPME provide very similar estimates, although again the harmonic RPME is the least sensitive to $\alpha_{min}$. In estimating the median, all RPME variants are identical because the median is never in the top bin.

Overall, any flavor of RPME can provide excellent estimates of the mean (98-99 percent reliability and 0-1 percent bias), very good estimates of the median (97 reliability and 1 percent bias) and good estimates of the Gini (83-87 percent reliability and 0-2 percent bias). However, the arithmetic RPME requires careful tuning of $\alpha_{min}$ to achieve good results, while the harmonic RPME is much more robust.

### *3.1.2 MGBE estimates*

Table 2 gives county estimates obtained by fitting various GB family distributions. Some distributions yield very poor estimates, which drives home the point that continuous distributions are only useful if they approximate the true distribution of income. Estimates from the Pareto distribution are the worst; Pareto-estimated Ginis have a bias of 20 percent and are only 23 percent reliable. Fortunately the fit statistics would never lead us to choose the Pareto distribution; in 98 percent of counties a $G^2$ test rejects the Pareto as the true distribution of income, and in 100 percent of counties there is some other distribution that fits better according to the AIC and BIC.

The best fitting distributions, according to the AIC and BIC, are the GB2, Dagum, gamma, and generalized gamma. The Weibull also fits well, although it rarely fits as well as the others. Even the best-fitting distributions are rejected by the $G^2$ test in at least three-quarters of counties. So they fit the data only approximately, although the approximation may be useful.

The bottom part of Table 2 summarizes the MGBE estimates that we get when we use the AIC or BIC to select the best-fitting of the 10 distributions. It also shows what happens when we average estimates across distributions using weights that are functions of the AIC and BIC. The average weight assigned each a model is not shown (to save space), but is very similar to the percentage of counties in which that model is chosen as best. The model-weighted averages are slightly more accurate than the model-selected estimates—a result that accords with statistical theory (Burnham & Anderson, 2004). The AIC and BIC produce almost identical estimates since they nearly always select the same distributions.

Looking at Table 2, one might get the impression that the 10-distribution MGBE is hardly worth the trouble since its estimates are barely better than those obtained from fitting the GB2 distribution alone. But this isn't quite right. As Table 3 shows, fitting just the GB2 distribution would fail to yield defined estimates in 4 percent of counties, whereas the MGBE is guaranteed to always produce a defined estimate. This implies that the accuracy statistics in Table 2 do not mean quite the same thing for the MGBE as they do for the GB2 distribution. In estimating the Gini, for example, the GB2's RMSE of 3 percent applies only to the 96 percent of counties where the GB2 produces a defined estimate. The MGBE achieves just as good a RMSE in those counties but also provides estimates for the remaining 4 percent. It is only when those 4 percent of counties are included that the MGBE has a RMSE of 4 percent.

Table 3 shows that several distributions besides the GB2 occasionally yield undefined estimates. The problem is worst for the log logistic distribution, which yields undefined estimates in more than half of counties. Table 3 also shows that several distributions occasionally have trouble with convergence. This is partly a software issue, but it is not unique to our software package in R. We also encountered similar convergence issues when we tried fitting GB2-family distributions to the same data in Stata and SAS. An issue is that it is tricky to define with binned data since the likelihood can be relatively flat near the maximum.

Table 3 also highlights another problem that arises when using the MGBE—the problem of runtime. The MGBE is much slower than the RPME, because the MGBE requires iteratively maximizing the likelihood of several distributions, some of which may have trouble with convergence. On our computer[10], it takes 32 hours to fit all 3,221 counties to all 10 GB distributions, but the RPME provides estimates for the same counties in a couple of seconds.

Table 3 also highlights several opportunities to substantially reduce runtime. The table shows that just three distributions—the Pareto 2, the beta 2, and the Singh-Maddala—account for 70 percent of the runtime in R.[11] It is not really necessary to run these three distributions since they get very little weight in the 10-model estimates. Some other distributions, notably the log logistic, are also unnecessary because they so often produce poor or undefined estimates.

We can get faster results by fitting just 2-4 models. For example, the Dagum, generalized gamma, and gamma account for just 11% of the total runtime in R, but get almost 70% of the weight in the county multimodel inferences. Running just these three distributions, we fit all 3,221 US counties in 3 hours using R. The bottom rows of Table 6 show that the resulting 3-model estimates are only slightly less accurate than the 10-model estimates. An even more radical solution would be to fit just the Weibull distribution, which consumes just 0.5 percent of the total runtime, always converges, and always provides a defined estimate. If only the Weibull distribution is fit, estimates for all 3,221 counties could be obtained in about 10 minutes. Even at 10 minutes, though the Weibull estimates would be much slower than the RPME, and no more accurate.

The scatterplots in Figure 5 help to visualize relative performance of the harmonic RPME and the MGBE. It is evident from the scatterplots that the MGBE mean and Gini estimates have a slight negative bias; that is, the points on the scatterplot tend to fall below the line. No such bias is evident in RPME estimates of the mean and Gini. On the other hand, RPME median estimates are slightly unreliable because they can only take values at the bin midpoint; this is evident from the stripes in the scatterplot.

### *3.1.3 Fewer bins*

As we wrote earlier, the RPME is a bin-consistent estimator that is at its best when the bins are numerous and narrow. With, say, 100 narrow bins, the RPME would be impossible for a parametric estimator like the MGBE to beat. Even with 16 bins, as we have seen, the RPME performs about as well as the MGBE.

How few bins can there be before the relative performance of the RPME starts to break down? Table 4 addresses this question by rebinning the county data. First we merge adjacent bins to reduce the number of bins from 16 to 8. Then we merge adjacent bins again to reduce the number of bins from 8 to 4.

With 8 bins, the performance of both estimators worsens, but the RPME remains about as accurate as the MGBE when estimating the mean and Gini. The RPME is worse than the MGBE at estimating the median; that was true with 16 bins as well, but with 8 bins the difference between the median estimates is more pronounced. In general, the RPME produces worse median estimates because it is discrete and assumes that the median can only be at a bin midpoint. This problem becomes more consequential when the bins are coarse.

With 4 bins, the RPME is almost as good as the MGBE when estimating the mean, but is notably worse in estimating the Gini and median. These 4-bin results will have limited interest to researchers using US Census data, which typically has at least 10 bins. But 4-bin results will be of interested to researchers who use data from the developing world, which sometimes has as few as 5 bins.

Which flavor of the RPME performs best with fewer bins? The answer is not clear-cut. With 8 bins, the arithmetic RPME is best at estimating the Gini, but the harmonic RPME is best at estimating the mean, and all RPME flavors are identical in estimating the median. With 4 bins the ranking of RMSE flavors depends on the criterion of accuracy (RMSE or reliability), but all flavors of RMSE are inferior to the MGBE.

## 3.2 States in 1980

Our evaluation continues with state-level summaries of family incomes reported on the long-form Census in 1980, which asked about family incomes received during 1979. For each US state and the District of Columbia, the Census reported the mean[12], median, and Gini of family income (US Census Bureau, 1982, 2010), and also summarized family incomes using 17 bins (National Historical Geographic Information System, 2005; US Census Bureau, 1982). The data give estimated bin counts for the population, and since the long-form Census was given to a 1-in-6 sample, we divide the population counts by 6 to estimate the sample counts. The long-form Census was given to a complex sample of households (Navarro & Griffin, 1990), but unfortunately the binned data provide no information about the sample design. We again analyze the data as if they came from a simple random sample, and this means that $p$ values etc. are only approximate (Rao & Scott, 1981).

Evans et al. (2004) previously analyzed these same data by fitting the Dagum distribution. We extend the analysis by fitting additional distributions from the GB family, combining the resulting estimates to obtain the MGBE, and comparing the MGBE to the RPME.

Table 5 and Figure 6 summarize estimates of the median, mean, and Gini of family income. Replicaing Evans et al.'s (2004) results, we find that excellent estimates can be obtained by fitting the Dagum distribution. The Dagum had a RMSE of 1 percent and a reliability of 98-99 percent in estimating the mean, median, and Gini.

Although the Dagum results leave little room for improvement, we find the estimates are just as good if we fit the Weibull, generalized gamma, or GB2 distributions. Other distributions are worse. Despite the good results obtained from some of the GB family distributions, the results show that all the GB distributions are only approximate fits to the binned data. In every state, a chi-square test rejected every fitted distribution as the true distribution of income.

Estimates obtained from the MGBE are practically the same as estimates obtained from either the Dagum or the GB2 alone. In fact, the MGBE gives 12 percent of its weight to the Dagum estimates and 88 percent to the GB2 estimates, with other distributions getting no weight.

The RPME is almost but not quite as accurate as the MGBE. In estimating the mean, the RPME is every bit as good as the MGBE, and in estimating the Gini, the RPME is just as reliable as the MGBE but has a small positive bias. In estimating the median, the RPME is unbiased but less reliable than the MGBE because the RPME median estimates can only take discrete values at the bin midpoints. The discreteness of the RPME median estimates is evident from the stripes in the scatterplot.

All the flavors of RMPE—arithmetic, geometric, harmonic, and median—produce very similar estimates. As in the county estimates, we used the arithmetic RPME with $\alpha_{min} = 2$ and the harmonic, geometric, and median RPMEs with $\alpha_{min} = 1$. These choices had no effect on the estimates since all the states had $\hat{\alpha} > 2$ anyway. In short, the robust features of the RPME, which are so necessary in fitting small county samples, were not needed for these large state samples.

## 3.3 States in 2005

Our evaluation continues with state-level summaries of household incomes reported to the American Community Survey (ACS) in 2005. For each state, the District of Columbia, and Puerto Rico, the Census published income distributions using the same 16 bins as those in Table 1 (US Census Bureau, n.d.). The data give estimated bin counts for the population, but since the ACS was given to a 1-in-40 sample, we divide the population counts by 40 to estimate the counts in the sample. Again, although the ACS is a complex sample, the binned data provide no information about the sample design, so we analyze the data as if they came from a simple random sample. This means that $p$ values etc. are only approximate (Rao & Scott, 1981).

What is especially valuable about the 2005 state data is that, for the first time, the Census published not just state Ginis, but also state Theils and MLDs (Hisnanick & Rogers, 2006),[13] providing an opportunity to evaluate how well our estimators reproduce these less common inequality statistics. Another valuable aspect of the 2005 data is that the distribution of family income changed substantially between 1980 and 2005, becoming much more unequal and stretching the upper tail (Western, Bloome, & Percheski, 2008). It is important to check whether the GB family distributions fit as well in 2005 as they did in 2008.

Table 6 and Figure 7 summarize state estimates of the mean, Gini, Theil, and MLD in 2005. As in 1980, the best fitting GB family distributions are the GB2 and, less often, the Dagum, although all the distributions are rejected by the goodness-of-fit test. In 1980, the rejection of the distributions' fit did not prevent them from providing good estimates, but in 2005 the estimates are worse. Perhaps that means the true income distribution is not approximated as well by the GB family.

For all estimators, the state estimates are worse in 2005 than they were in 1980, and worse for the Theil and MLD than for the Gini. The 2005 RPME estimates are a little better than the MGBE estimates—which is a reversal of the findings for 1980—but both estimators have similar problems. In estimating the mean, both the RPME and the MGBE are excellent. In estimating the Gini, the MGBE has a small negative bias but is highly reliable, while the RPME is a little less reliable but has less bias and a smaller RMSE.

In estimating the Theil and MLD both estimators have large negative biases. Why are the relative biases so much worse for the Theil and MLD than they are for the Gini? We can get some intuition for the answer by pointing out that, empirically, the true Theil and MLD can be very well approximated by linear functions of the true Gini. The following are the least-squares lines for the states in 2005:

$$\begin{aligned} Theil &\approx -.46 + 1.86\,Gini \quad (R^2 = .96) \\ MLD &\approx -.78 + 2.83\,Gini \quad (R^2 = .92) \end{aligned} \tag{8}$$

Consider first the equation for the Theil. The fact that the slope is 1.86 suggests that the bias of the Theil estimate will be approximately 1.86 times that of the Gini estimate. However, the fact that the intercept is negative means that the true Theil is less than 1.86 times the true Gini. In relative terms, then, the bias will be a larger percentage of the Theil than it is of the Gini. A similar argument holds for the MLD.

The different flavors of RPME performed similarly. As in the county estimates, we used the arithmetic RPME with $\alpha_{min} = 2$ and the harmonic, geometric, and median RPMEs with $\alpha_{min} = 1$. These choices had no effect on the estimates since all the states (except for the District of Columbia) had $\hat{\alpha} > 2$ anyway. Again, the robust features of the RPME, which are so necessary in fitting small county samples, were not needed for these large state samples.

## 3.4 National data from 1970 to 2006-09

Up this point we have found that the RPME and MGBE have comparable accuracy in the common situation where there are at least 8 bins. It is tempting to conclude that it doesn't matter which estimator you use. That conclusion might be valid for a cross-sectional analysis.

In a longitudinal analysis, however, the choice of estimator can matter. While the errors in Gini estimates are small compared to cross-sectional variation in the Gini, the errors are not trivial compared to the trends in the Gini over time.

To illustrate this point, we use national data on family incomes in every decade from 1970 to 2006-09. Binned data are available from the long-form Census from 1970, 1980, 1990, and 2000, and from the American Community Survey in 2006-09 (adjusted to 2009 dollars). There are 25 bins in 1990, and 15-17 bins in other years. In each year, we estimate the Gini using the MGBE and RPME, and compare the resulting estimates to the "true" Gini calculated by the Census using unbinned data from the Current Population Survey.

As in previous analyses, we use the arithmetic RPME with $\alpha_{min} = 2$ and the harmonic, geometric, and median RPMEs with $\alpha_{min} = 1$. Our choice of $\alpha_{min}$ has no effect on the estimates since in every year the estimate $\hat{\alpha}$ is greater than 2 anyway. Our MGBE estimates select the best of 10 distributions according to the AIC. In 2006-09 the best distribution was the GB2; in all other years it was the Dagum.

Table 7 gives true and estimated values for the national Gini in every year. Clearly the MGBE estimates do a better job than the RPME in capturing the true uptrend in the Gini. The true Gini increases in every decade, and so does the MGBE estimate, but the RPME estimates do not. To the contrary, every flavor of RPME estimates that the Gini decreased between 1970 and 1980, and did not increase between 2000 and 2006-09. Although every estimator captures the big increases in the Gini in between 1980 and 2000, the smaller increases before and after can be swamped by estimation error. The estimation errors are not huge, 10% at the most, but because the errors change from year to year they can obscure the underlying trends.

Overall the MGBE has a smaller RMSE, but this is entirely because of its better performance in 1970. Excluding that one year, the MGBE and RPME have very similar RMSEs. The real advantage of the MGBE is that it does a better job of capturing the trend.

The fact that the MGBE captures the trend in these particular data does not mean that it would always capture trends better than the RPME. Before conducting a detailed analysis, it is useful to consult a few known "true" values, as we do in Table 7, to see which estimator best captures the known trends in a particular dataset.

## 4. DISCUSSION

Although the RPME and MGBE often produce very good estimates, they do leave room for improvement, particularly in estimating the Theil and MLD.

### 4.1 Ideas for improvement

While the RPME is bin-consistent, its major weakness is that it treats income as a discrete variable. Alternatives are possible that treat income as continuous while preserving the property of bin-consistency. The simplest idea is to spread each bin's cases evenly across their bin, so that each bin has its own uniform density—except for the top bin, which is given a Pareto density. This approach is simple but tends to exaggerate inequality because the fitted uniform distributions are too flat in the tails (Cloutier, 1988).

The idea of fitting each bin to its own density can be improved. One improvement is to fit each of the right bins to its own downward sloping density (Jargowsky, 1995) and fit each of the left bins to its own upward sloping density. In addition, alternatives to the Pareto density may be considered for the top bin. Unfortunately, distributions that fit each bin to its own density are not continuous at the bin boundaries. They also are not identified because they have as many parameters as bins. An attempt to solve both problems would be to fit each bin to its own density, but constrain the densities to be continuous at the bin boundaries.

While the MGBE treats income as continuous, its major weakness is that it uses only unimodal GB family densities. These densities are usually rejected as the true distribution of income, and in some data they may be poor approximations. The multimodel approach could be improved by adding densities from outside the GB family, especially bimodal densities which the GB family cannot mimic.

These suggested improvements converge on the idea of fitting a density that is as smooth as the GB family densities but as flexible and assumption-free as the RPME. The idea is attractive, but it may be difficult to achieve in practice. A very flexible continuous distribution would have many parameters, and it gets hard to achieve identification or convergence as the number of parameters approaches the number of bins. In our experience, it can be difficult getting the 4-parameter GB2 distribution to converge; a distribution with more parameters would probably be even more challenging. It is helpful to parameterize distributions so that each parameter captures a distinct aspect of the distribution's shape (Stasinopoulos & Rigby, 2013), but this can be hard to achieve when there are many parammeters.

There have been two attempts to fit binned incomes to flexible non-GB distributions that can be either unimodal or multimodal. One attempt is the logspline method, which fits binned data to a density whose logarithm is a polynomial spline (Kooperberg & Stone, 1992). A second attempt is kernel density estimation, which requires bin means as well as bin counts (e.g., Sala-i-Martin, 2006). While the ideas behind these methods are appealing, unfortunately neither has performed well in practice. Both methods can display substantial biases when fit to real or simulated income distributions, and both methods perform worse than simply fitting the GB2 distribution (Minoiu & Reddy, 2012; von Hippel et al., 2012).[14]

## 4.2 Using auxiliary statistics

Further improvements are possible if the data offer statistics beyond the bin counts. As mentioned previously, some data give the bin means as well as the bin counts. These can be used to improve the RPME since we can simply replace the bin midpoints with the bin means and there is no need for a Pareto assumption in the top bin. Theoretically, the bin means can also be used to improve the estimation of GB family distributions using a generalized method of moments (GMM) estimator. However, the GMM estimator is difficult to optimize and can break down when the second moment does not exist or the estimates get close to the edge of the parameter space (Hajargasht, Griffiths, Brice, Rao, & Chotikapanich, 2012). Bin means are also needed for kernel density estimation; however, the results obtained by applying kernel density estimation to binned data have been disappointing, as previously discussed (Minoiu & Reddy, 2012). Bin means are also needed for interpolation methods which fit a simple density within each bin, subject to the constraint that the fitted density must reproduce the known bin mean (Cowell & Mehta, 1982).

While bin means are only available occasionally, it is quite common for the Census to provide the grand mean. Then binned-data estimates can be improved by constraining the estimates to match the grand mean. The improvement will be very small, however, since even without constraints RPME and MGBE already match the mean with high accuracy. In addition, the constraints on the GB family distributions will be a nonlinear function of several parameters and will be difficult to implement in software.[15]

It is also common for data to provide the grand median. Having the grand median is equivalent to having an extra bin, since the investigator can split one of the bins at the median and easily figure out the counts for the split bins. Adding an extra bin will improve estimates from either the RPME or MGBE, but the improvement will be small if there are already many bins or the median is close to a bin boundary.

Methods for improving the estimation of GB parameters with auxiliary statistics will not be fruitful if the true income distribution is not well approximated by a GB family distribution. For example, if the true income distribution is bimodal, then fitting a unimodal GB distribution is a procrustean effort that will result in bias even if auxiliary statistics are available.

# 5. CONCLUSION

This is the broadest evaluation of binned-data estimators that has ever been carried out. We evaluated two of the most popular estimators and found that neither was robust in small samples. We developed robust versions of the estimators which we called the robust Pareto midpoint estimator (RPME) and the multimodel GB estimator (MGBE). We have implemented these estimators in the *rpme* and *mgbe* commands for Stata and the *binequality* package for R (Duan & von Hippel, 2016; Scarpino et al., 2014; von Hippel & Powers, 2014).

The MGBE is typically more accurate if there are less than 8 bins, but with more bins the two methods typically produce very similar and very good estimates of the mean and Gini. It is likely that both methods also produce good estimates of the CV because the CV and Gini are closely related (Hosking, 1990; Milanovic, 1997). However, both methods can produce poor estimates of the Theil and MLD.

Although errors in estimating the Gini are typically small, they can be large enough to affect estimated trends in the Gini over time. When using the methods to estimate trends, it is helpful to use reference statistics to check which method is giving better estimates.

An advantage of the RPME is that it runs much faster than the iterative MGBE. This can be an important consideration when one is working with hundreds or thousands of binned datasets, each of which represents a different county, school district, or neighborhood.

Our discussion suggests potential improvements to both methods. Some of the improvements would not be easy but, if attempted, could be tested using the relatively new data in this paper.

## Endnotes

[1] There are two justifications for treating the bottom of the lower bin as zero. First, the distributions that are commonly fit to incomes only support non-negative values. Second, according to IPUMS data only 1.6% of bottom-bin households, or 0.05% of all households, have negative incomes, so rounding those incomes up to zero has negligible effects on the estimates. In our results, for example, we will not find that treating bottom bin as bounded at 0 results in overestimates of the mean or median, even in poor communities.
[2] For Maricao the average width of the populated bins is $6,818, and the estimated SD is $\hat{\sigma}_{ME} = \$14{,}307$. The ratio of these quantities is about 1/2.
[3] The Pareto assumption is arbitrary. We worked briefly on an alternative that fit the top two bins to a truncated Weibull distribution, but the resulting formulas were more complicated and did not produce better results.
[4] We tried West’s alternative estimator in our data and confirmed that it produced results worse than $\hat{\alpha}$.
[5] A more complete family tree would include the 4-parameter GB1 distribution and a 5-parameter GB distribution that nests both the GB1 and the GB2 (McDonald & Xu, 1995). We have excluded the GB and GB1 because (1) they rarely fit income distributions better than the GB2 (Bandourian, McDonald, & Turley, 2002); and (2) they are not implemented in available software. In fact, the GB and GB1 would be difficult to implement because their parameters include a lower bound which would be difficult to estimate from binned data (Robert Rigby, personal communication).
[6] Stasinopouls et al. (2008) allow $\sigma$ to be negative, but we constrained it to be positive by using a log link—i.e., by estimating $\ln(\sigma)$.
[7] It is possible to sample from a distribution with an undefined mean and calculate an average from the sample. But this sample average is meaningless and, if calculated, can be arbitrarily large and variable.

[8] Notice that the fit of a model can only be tested if *df* is at least 1. For example, if less than $B$=5 bins are populated, it will not be possible to test a distribution with more than $k$=3 parameters.
[9] The authors did not remark on these rejections, but you can see them by comparing the reported $X^2$ values to the critical values from a chi-square distribution.
[10] We used an Intel i7 Quad Core MacBook Pro laptop which has a 2.0GHz clock speed, 16GB of RAM, and a 500GB hard drive.
[11] The Stata implementation is different. Using our *mgbe* command for Stata, all the distributions run equally quickly.
[12] The Census actually reported total ("aggregate") family income. We calculated mean family income by dividing total income by the number of families.
[13] The state Theils and MLDs are "unofficial" statistics published in a report by Census employees (Hisnanick & Rogers, 2006), which was not vetted as thoroughly as an official Census report. Initially, typos in the unofficial report made us concerned about the accuracy of the unofficial estimates. Later, however, we noticed that the national Theil and MLD in the unofficial report agreed closely with the national Theil and MLD given in an official Census report based on the current population survey (DeNavas-Walt, Proctor, & Smith, 2013)..
[14] It is not clear whether the observed bias of logspline estimates is due to the logspline method itself, or due to the way that it is implemented in legacy software (namely the *oldlogspline* package for R).
[15]The *gamlss* package for R, which we used to implement the MGBE, has no provision for multiparameter constraints (Stasinopoulos, Rigby, & Akantzilioutou, 2008).

# TABLES

Table 1. Distribution of household income in two U.S. counties

| | | | Households in county | |
|---|---|---|---|---|
| Bin | Min | Max | Maricao | Nantucket |
| 1 | | $10,000 | 781 | 165 |
| 2 | $10,000 | $15,000 | 245 | 109 |
| 3 | $15,000 | $20,000 | 140 | 67 |
| 4 | $20,000 | $25,000 | 156 | 147 |
| 5 | $25,000 | $30,000 | 85 | 114 |
| 6 | $30,000 | $35,000 | 60 | 91 |
| 7 | $35,000 | $40,000 | 37 | 148 |
| 8 | $40,000 | $45,000 | 61 | 44 |
| 9 | $45,000 | $50,000 | 9 | 121 |
| 10 | $50,000 | $60,000 | 57 | 159 |
| 11 | $60,000 | $75,000 | 19 | 358 |
| 12 | $75,000 | $100,000 | 0 | 625 |
| 13 | $100,000 | $125,000 | 0 | 338 |
| 14 | $125,000 | $150,000 | 0 | 416 |
| 15 | $150,000 | $200,000 | 0 | 200 |
| 16 | $200,000 | | 0 | 521 |
| Total | | | 1,650 | 3,623 |

*Note*. The number of households in each bin is estimated from a 1-in-8 sample of households who took the American Community Survey in 2006-10. Incomes reported in the survey have been inflated to 2010 dollars.

Table 2. County estimates in 2006-10

| | | Estimand | | | | | | | | | % selected | | |
|---|---|---|---|---|---|---|---|---|---|---|---|---|---|
| | | Median | | | Mean | | | Gini | | | | | |
| Estimator | Type | % bias | % RMSE | % reliable | % bias | % RMSE | % reliable | % bias | % RMSE | % reliable | by AIC | by BIC | Fit rejected |
| RPME | Arithmetic | 1 | 5 | 97 | 1 | 3 | 99 | 0 | 3 | 87 | | | |
| | Geometric | 1 | 5 | 97 | 0 | 3 | 98 | -1 | 4 | 83 | | | |
| | Median | 1 | 5 | 97 | -1 | 3 | 99 | -2 | 4 | 84 | | | |
| | Harmonic | 1 | 5 | 97 | 0 | 3 | 99 | -2 | 4 | 85 | | | |
| GB family distributions | Weibull | 2 | 4 | 98 | -2 | 4 | 98 | -4 | 5 | 83 | 6 | 7 | 85 |
| | Log logistic | -5 | 6 | 99 | 12 | 13 | 97 | 13 | 14 | 47 | 1 | 1 | 90 |
| | Pareto 2 | -16 | 17 | 91 | -2 | 6 | 93 | 20 | 22 | 23 | 0 | 0 | 98 |
| | Gamma | 0 | 4 | 98 | -2 | 4 | 98 | -4 | 5 | 81 | 21 | 24 | 81 |
| | Log normal | -9 | 9 | 98 | 3 | 5 | 98 | 6 | 8 | 73 | 3 | 3 | 91 |
| | Dagum | 1 | 3 | 99 | 1 | 3 | 99 | 0 | 4 | 83 | 28 | 29 | 77 |
| | Singh-Maddala | -2 | 4 | 97 | -1 | 3 | 97 | -1 | 6 | 77 | 7 | 8 | 80 |
| | Beta 2 | -5 | 6 | 98 | 0 | 4 | 98 | 1 | 5 | 73 | 1 | 1 | 86 |
| | Generalized gamma | -1 | 3 | 99 | -2 | 3 | 99 | -4 | 5 | 86 | 19 | 16 | 77 |
| | GB2 | 0 | 3 | 98 | -1 | 3 | 97 | -1 | 3 | 87 | 15 | 11 | 76 |
| MGBE, 10 models | AIC-selected | 0 | 3 | 99 | -1 | 3 | 99 | -2 | 4 | 87 | | | |
| | BIC-selected | 0 | 3 | 99 | -1 | 3 | 99 | -2 | 4 | 87 | | | |
| | AIC-averaged | 0 | 3 | 99 | -1 | 3 | 99 | -2 | 4 | 88 | | | |
| | BIC-averaged | 0 | 3 | 99 | -1 | 3 | 99 | -2 | 4 | 88 | | | |
| MGBE, 3 models | AIC-selected | 0 | 3 | 99 | -1 | 3 | 99 | -2 | 4 | 86 | | | |
| | BIC-selected | 0 | 3 | 99 | -1 | 3 | 99 | -3 | 4 | 86 | | | |
| | AIC-averaged | 0 | 3 | 99 | -1 | 3 | 99 | -2 | 4 | 87 | | | |
| | BIC-averaged | 0 | 3 | 99 | -1 | 3 | 99 | -3 | 4 | 87 | | | |

*Note*. The arithmetic RPME was used with $\alpha_{min} = 2$; the harmonic, geometric, and median RPMEs were used with $\alpha_{min} = 1$. Rejections of fit were based on a likelihood ratio chi-square test with a significance level of .05.

Table 3. Estimation problems when fitting GB family distributions

| Parameters | Model | % undefined estimates | % non-convergence | % total runtime |
|---|---|---|---|---|
| 2 | Weibull | 0.0 | 0 | 0.5 |
| | Log logistic | 53.1 | 0 | 1.3 |
| | Pareto 2 | 1.3 | 1 | 34.0 |
| | Gamma | 0.0 | 0 | 0.4 |
| | Log normal | 0.0 | 0 | 0.5 |
| 3 | Dagum | 4.2 | 0 | 8.8 |
| | Singh-Maddala | 0.4 | 2 | 16.2 |
| | Beta 2 | 0.0 | 11 | 20.1 |
| | Generalized gamma | 0.6 | 0 | 1.7 |
| 4 | GB2 | 4.4 | 1 | 16.5 |

Table 4. Accuracy of county estimates, with fewer bins

| | | | Estimand | | | | | | | | |
|---|---|---|---|---|---|---|---|---|---|---|---|
| | | | Median | | | Mean | | | Gini | | |
| Bins | Estimator | Type | % bias | % RMSE | % reliable | % bias | % RMSE | % reliable | % bias | % RMSE | % reliable |
| 8 | RPME | Arithmetic | 2 | 11 | 89 | 2 | 4 | 98 | -1 | 4 | 78 |
| | | Geometric | 2 | 11 | 89 | 1 | 4 | 97 | -2 | 5 | 78 |
| | | Harmonic | 2 | 11 | 89 | 0 | 3 | 98 | -4 | 5 | 76 |
| | | Median | 2 | 11 | 89 | -1 | 3 | 98 | -5 | 6 | 74 |
| | MGBE | 10 distributions, best AIC | 0 | 3 | 99 | -4 | 5 | 98 | -4 | 6 | 79 |
| 4 | RPME | Arithmetic | 9 | 28 | 64 | 12 | 13 | 95 | 0 | 7 | 38 |
| | | Geometric | 9 | 29 | 62 | 11 | 14 | 94 | -2 | 7 | 44 |
| | | Harmonic | 9 | 28 | 64 | 4 | 6 | 96 | -7 | 9 | 53 |
| | | Median | 9 | 28 | 64 | 3 | 6 | 96 | -9 | 10 | 55 |
| | MGBE | 10 distributions, best AIC | 0 | 3 | 99 | -5 | 6 | 96 | -6 | 8 | 65 |

Note. The arithmetic RPME uses $\alpha_{min} = 2$, while the geometric, harmonic, and median RPME use $\alpha_{min} = 1$.

Table 5. Accuracy of state estimates in 1980

| | | Estimand | | | | | | | | | | | |
|---|---|---|---|---|---|---|---|---|---|---|---|---|---|
| | | Median | | | Mean | | | Gini | | | % selected | | |
| Estimator | Flavor | % bias | % RMSE | % reliable | % bias | % RMSE | % reliable | % bias | % RMSE | % reliable | by AIC | by BIC | % fit rejected |
| RPME | Arithmetic | 0 | 4 | 94 | 1 | 1 | 99.7 | 3 | 3 | 97.2 | | | |
| | Geometric | 0 | 4 | 94 | 1 | 1 | 99.7 | 2 | 2 | 97.2 | | | |
| | Harmonic | 0 | 4 | 94 | 0 | 1 | 99.7 | 2 | 2 | 97.2 | | | |
| | Median | 0 | 4 | 94 | 0 | 1 | 99.7 | 1 | 1 | 97.2 | | | |
| GB family distributions | Weibull | -1 | 1 | 98 | -1 | 2 | 99.6 | 0 | 1 | 99.6 | 0 | 0 | 100 |
| | Log logistic | -6 | 6 | 99 | 11 | 12 | 97.0 | 16 | 16 | 84.3 | 0 | 0 | 100 |
| | Pareto 2 | -23 | 23 | 91 | -2 | 6 | 86.1 | 39 | 40 | 29.6 | 0 | 0 | 100 |
| | Gamma | -3 | 4 | 98 | -1 | 1 | 99.6 | 1 | 2 | 95.0 | 0 | 0 | 100 |
| | Log normal | -11 | 11 | 98 | 4 | 4 | 98.9 | 14 | 14 | 83.8 | 0 | 0 | 100 |
| | Dagum | -1 | 1 | 99 | -1 | 1 | 99.7 | 0 | 1 | 98.1 | 12 | 12 | 100 |
| - | Singh-Maddala | -3 | 4 | 97 | -1 | 2 | 99.4 | 1 | 4 | 64.2 | 0 | 0 | 100 |
| | Beta 2 | -6 | 6 | 98 | -1 | 1 | 99.7 | 4 | 4 | 95.4 | 0 | 0 | 100 |
| | Generalized gamma | -3 | 3 | 99 | -1 | 1 | 99.7 | 0 | 0 | 99.8 | 0 | 0 | 100 |
| - | GB2 | 0 | 0 | 98 | 0 | 1 | 99.9 | 0 | 1 | 97.7 | 88 | 88 | 100 |
| MGBE (10 models) | AIC-selected | 0 | 0 | 99 | 0 | 1 | 99.9 | 0 | 1 | 97.8 | | | |
| | BIC-selected | 0 | 0 | 99 | 0 | 1 | 99.9 | 0 | 1 | 97.8 | | | |
| | AIC-weighted | 0 | 0 | 99 | 0 | 1 | 99.9 | 0 | 1 | 97.7 | | | |
| | BIC-weighted | 0 | 0 | 99 | 0 | 1 | 99.9 | 0 | 1 | 97.7 | | | |

Table 6. Accuracy of state estimates in 2005

| | | Estimand | | | | | | | | | | | | | | |
|---|---|---|---|---|---|---|---|---|---|---|---|---|---|---|---|---|
| | | Mean | | | Gini | | | Theil | | | MLD | | | % selected | | |
| | | % bias | % RMSE | % reliable | % bias | % RMSE | % reliable | % bias | % RMSE | % reliable | % bias | % RMSE | % reliable | by AIC | by BIC | % fit rejected |
| RPME- | Arithmetic | 3 | 3 | 99 | -1 | 1 | 87 | -6 | 7 | 72 | -18 | 18 | 88 | | | |
| | Geometric | 1 | 2 | 99 | -2 | 3 | 87 | -11 | 11 | 72 | -20 | 20 | 88 | | | |
| | Harmonic | 1 | 1 | 99 | -3 | 3 | 87 | -13 | 14 | 72 | -21 | 21 | 88 | | | |
| | Median | 0 | 1 | 99 | -4 | 4 | 87 | -15 | 16 | 72 | -22 | 22 | 88 | | | |
| GB family distributions- | Weibull | -2 | 3 | 99 | -6 | 6 | 89 | -23 | 23 | 89 | -24 | 24 | 94 | 0 | 0 | 100 |
| | Log logistic | 10 | 10 | 99 | 8 | 8 | 69 | 11 | 13 | 73 | -14 | 15 | 73 | 0 | 0 | 100 |
| | Pareto 2 | -6 | 8 | 90 | 13 | 14 | 53 | 14 | 18 | 76 | 17 | 23 | 76 | 0 | 0 | 100 |
| | Gamma | -2 | 3 | 99 | -6 | 6 | 91 | -24 | 24 | 90 | -28 | 28 | 96 | 0 | 0 | 100 |
| | Log normal | 3 | 3 | 99.5 | 3 | 4 | 88 | -3 | 6 | 90 | -23 | 23 | 93 | 0 | 0 | 100 |
| | Dagum | -1 | 1 | 99.7 | -4 | 4 | 96 | -17 | 18 | 94 | -24 | 24 | 96 | 12 | 12 | 100 |
| | Singh-Maddala | -2 | 2 | 99 | -5 | 5 | 93 | -20 | 21 | 91 | -28 | 28 | 96 | 0 | 0 | 100 |
| | Beta 2 | -2 | 2 | 99 | -4 | 4 | 95 | -18 | 19 | 92 | -28 | 28 | 96 | 0 | 0 | 100 |
| | Gen. gamma | -2 | 3 | 99 | -5 | 6 | 92 | -22 | 23 | 90 | -29 | 29 | 96 | 0 | 0 | 100 |
| | GB2 | -2 | 2 | 99.6 | -5 | 5 | 96 | -19 | 20 | 93 | -25 | 26 | 96 | 88 | 88 | 100 |
| MGBE (10 models) | AIC-selected | -2 | 2 | 99.6 | -5 | 5 | 96 | -19 | 19 | 93 | -25 | 25 | 96 | | | |
| | BIC-selected | -2 | 2 | 99.6 | -5 | 5 | 96 | -19 | 19 | 93 | -25 | 25 | 96 | | | |
| | AIC-weighted | -2 | 2 | 99.6 | -5 | 5 | 96 | -19 | 19 | 93 | -25 | 25 | 96 | | | |
| | BIC- weighted | -2 | 2 | 99.6 | -5 | 5 | 95 | -19 | 19 | 93 | -25 | 25 | 96 | | | |

Table 7. National estimates for the Gini of family income

| | | | | RPME | | | | | | | |
|---|---|---|---|---|---|---|---|---|---|---|---|
| | | MGBE | | Arithmetic | | Geometric | | Median | | Harmonic | |
| Decade | True | Estimate | Error | Estimate | Error | Estimate | Error | Estimate | Error | Estimate | Error |
| 1970 | .349 | .353 | 1% | .385 | 10% | .382 | 9% | .378 | 8% | .380 | 9% |
| 1980 | .365 | .362 | -1% | .380 | 4% | .376 | 3% | .371 | 2% | .374 | 2% |
| 1990 | .401 | .392 | -2% | .419 | 4% | .410 | 2% | .403 | 0% | .406 | 1% |
| 2000 | .429 | .404 | -6% | .431 | 0% | .420 | -2% | .410 | -4% | .414 | -3% |
| 2006-09 | .439 | .411 | -6% | .431 | -2% | .419 | -5% | .407 | -7% | .413 | -6% |
| Bias | | | -3% | | 4% | | 2% | | 0% | | 1% |
| RMSE | | | 4% | | 6% | | 6% | | 6% | | 6% |

## FIGURES

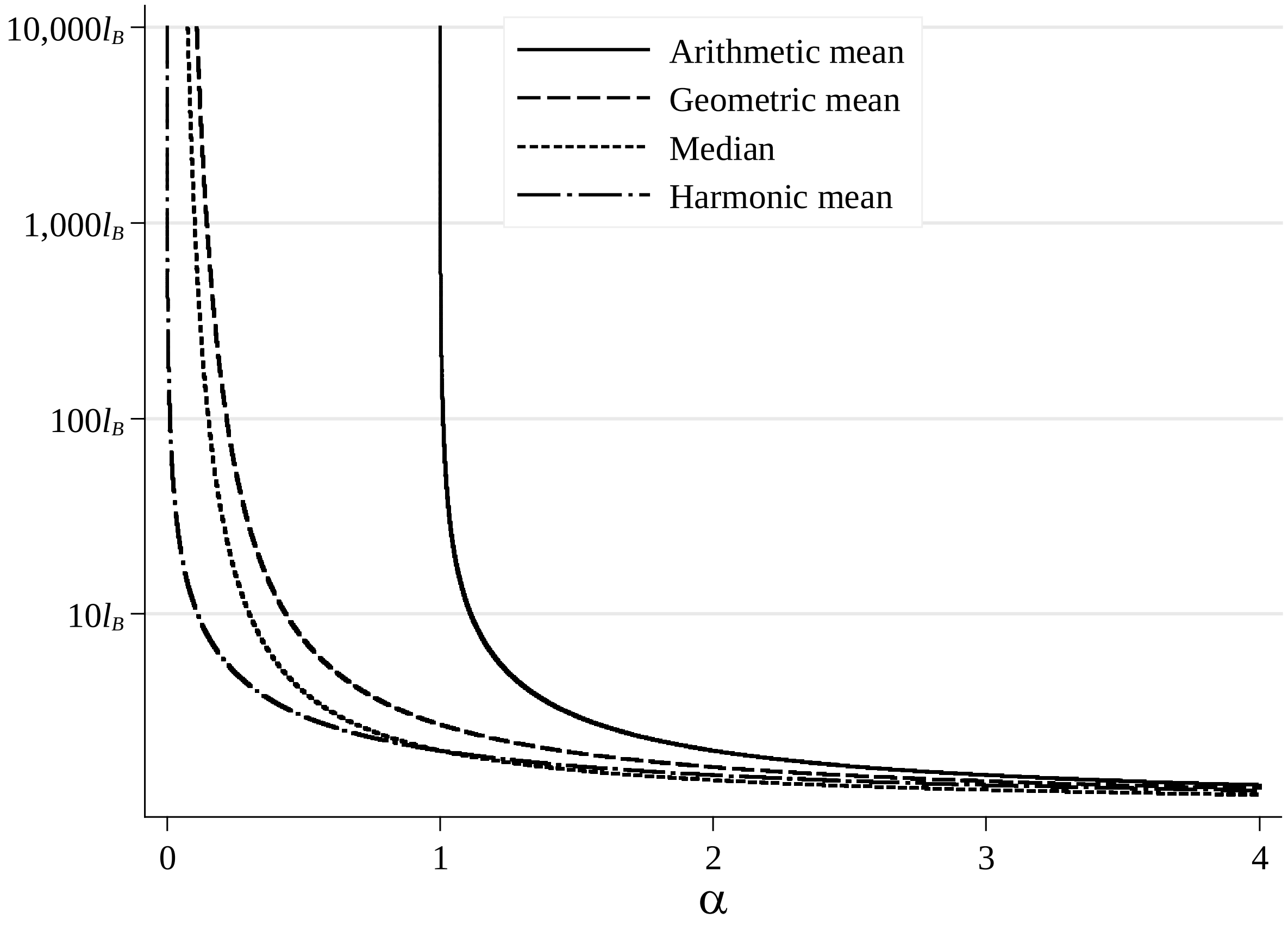

Figure 1. The median and three types of mean graphed as functions of the Pareto shape parameter $\alpha$. (*Note*. The verical axis is on a log scale.)

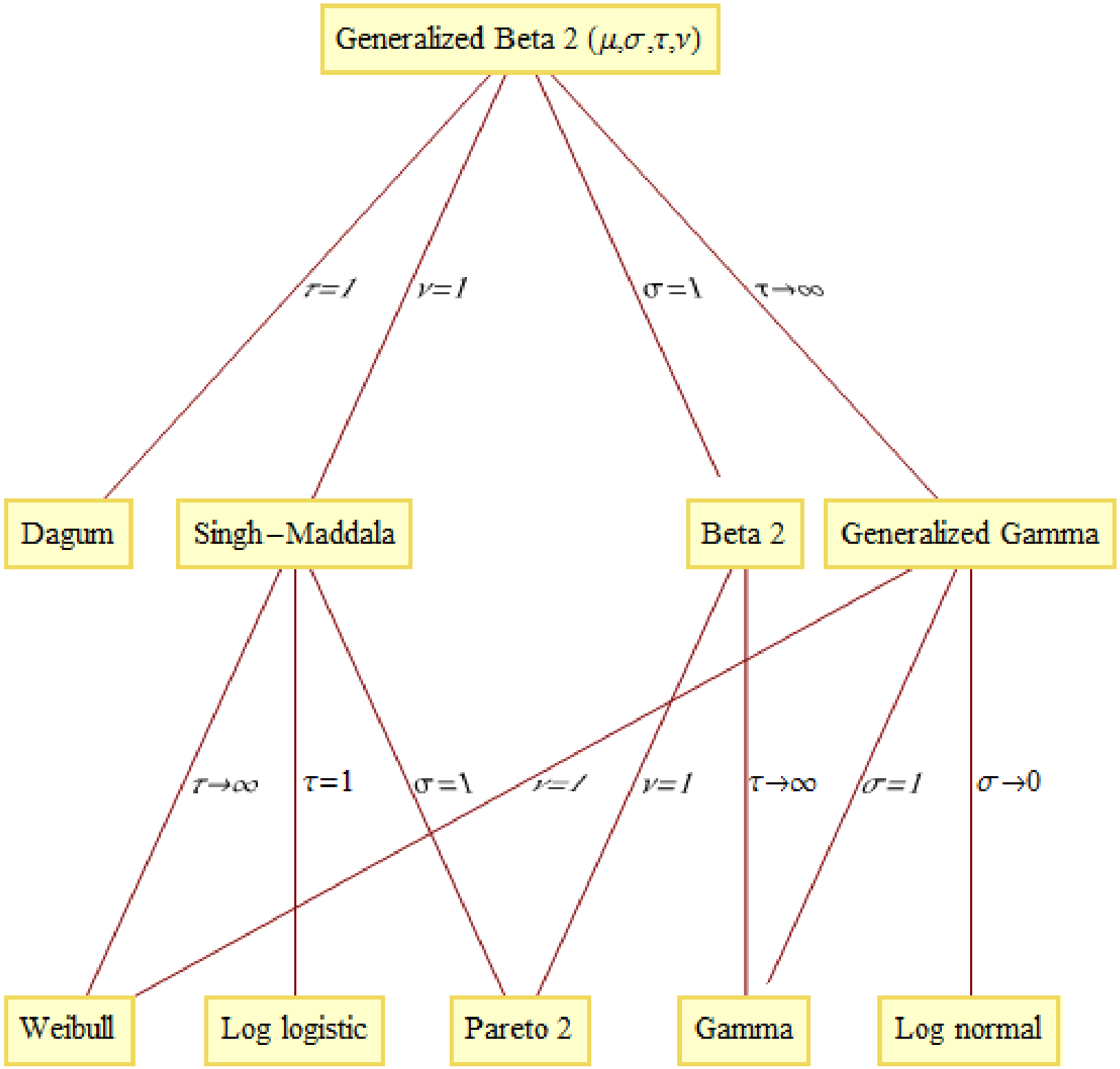

Figure 2. Part of the generalized beta family tree.

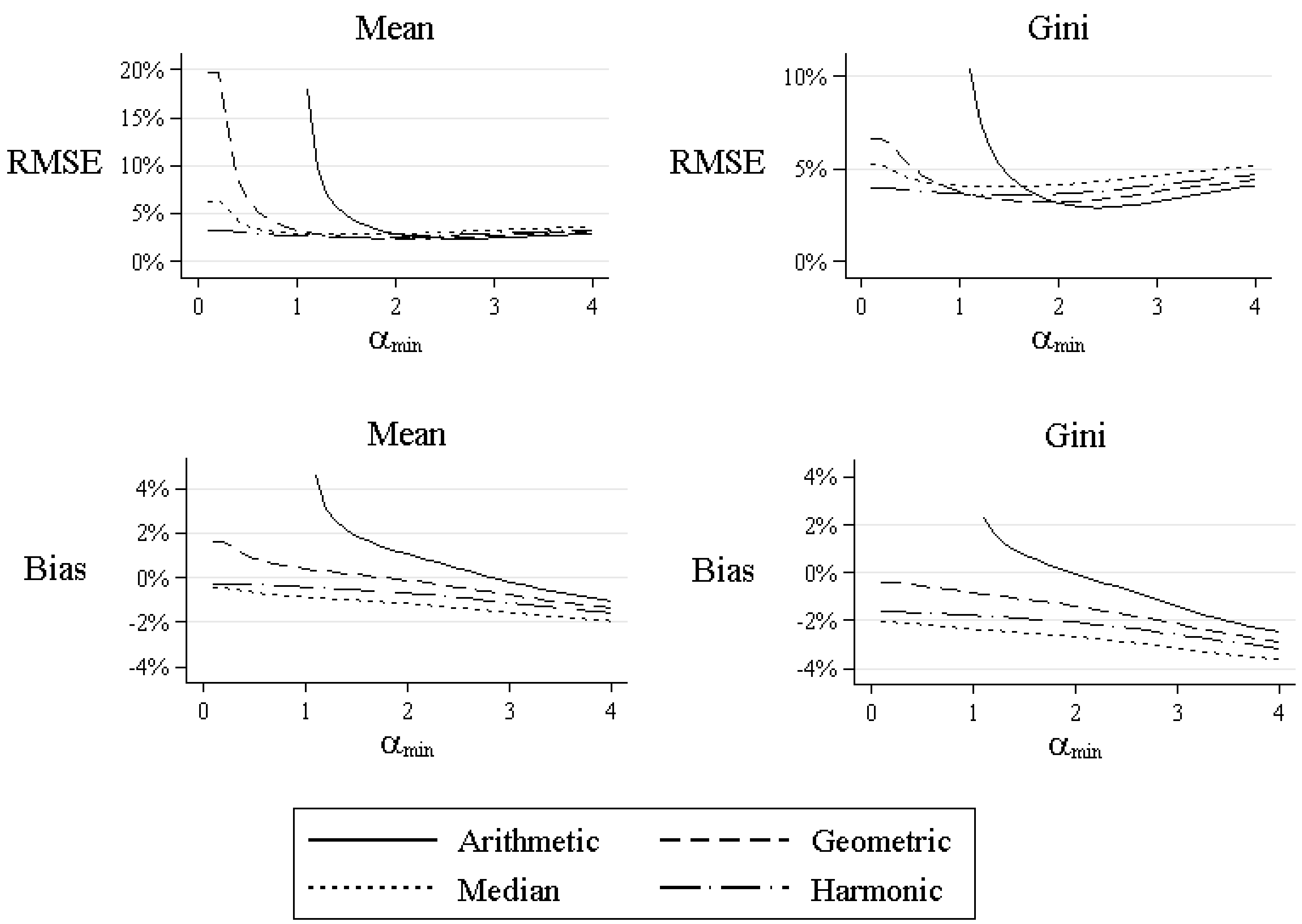

Figure 3. RMSE and bias for variants of the RPME in estimating the county means and Gini.

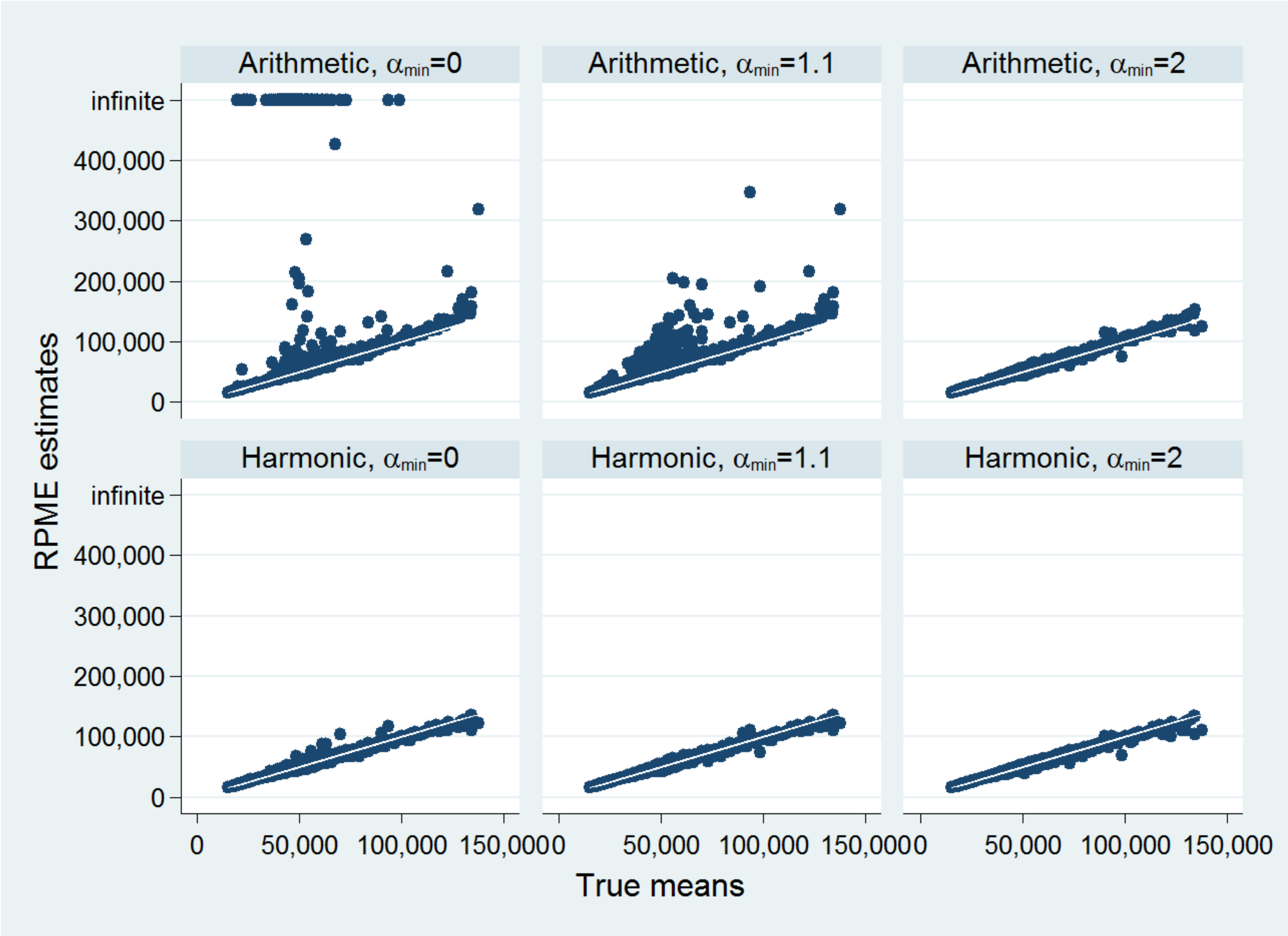

Figure 4. True vs. estimated county means provided by different variants of the RPME. An *X*=*Y* reference line shows what perfect estimation would look like.

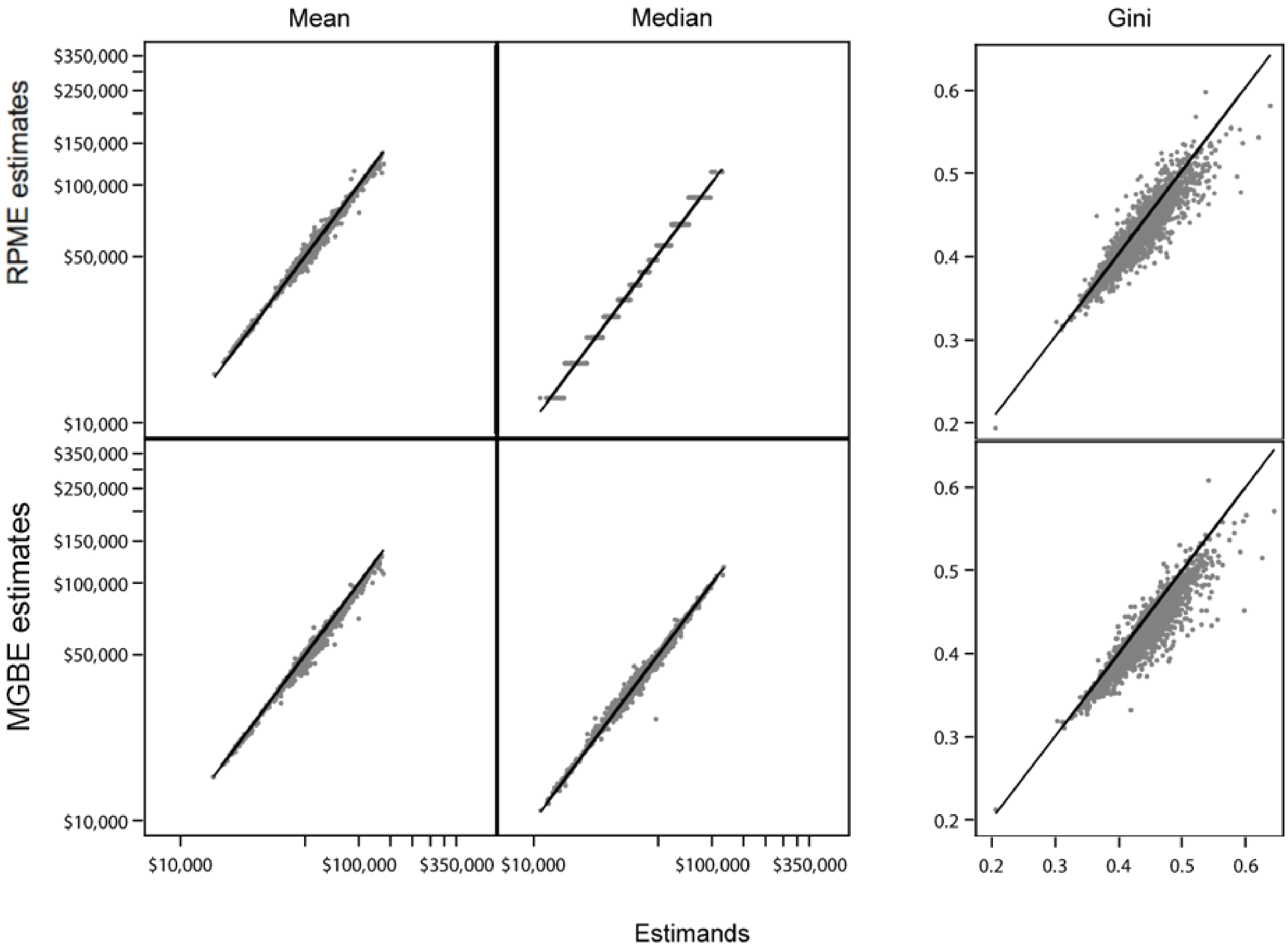

Figure 5. True vs. estimated county means, medians, and Ginis means for the harmonic RPME (with $\alpha_{min} = 1$) and the MGBE (using the AIC to select the best of 10 models.) An *X*=*Y* reference line shows what perfect estimation would look like..

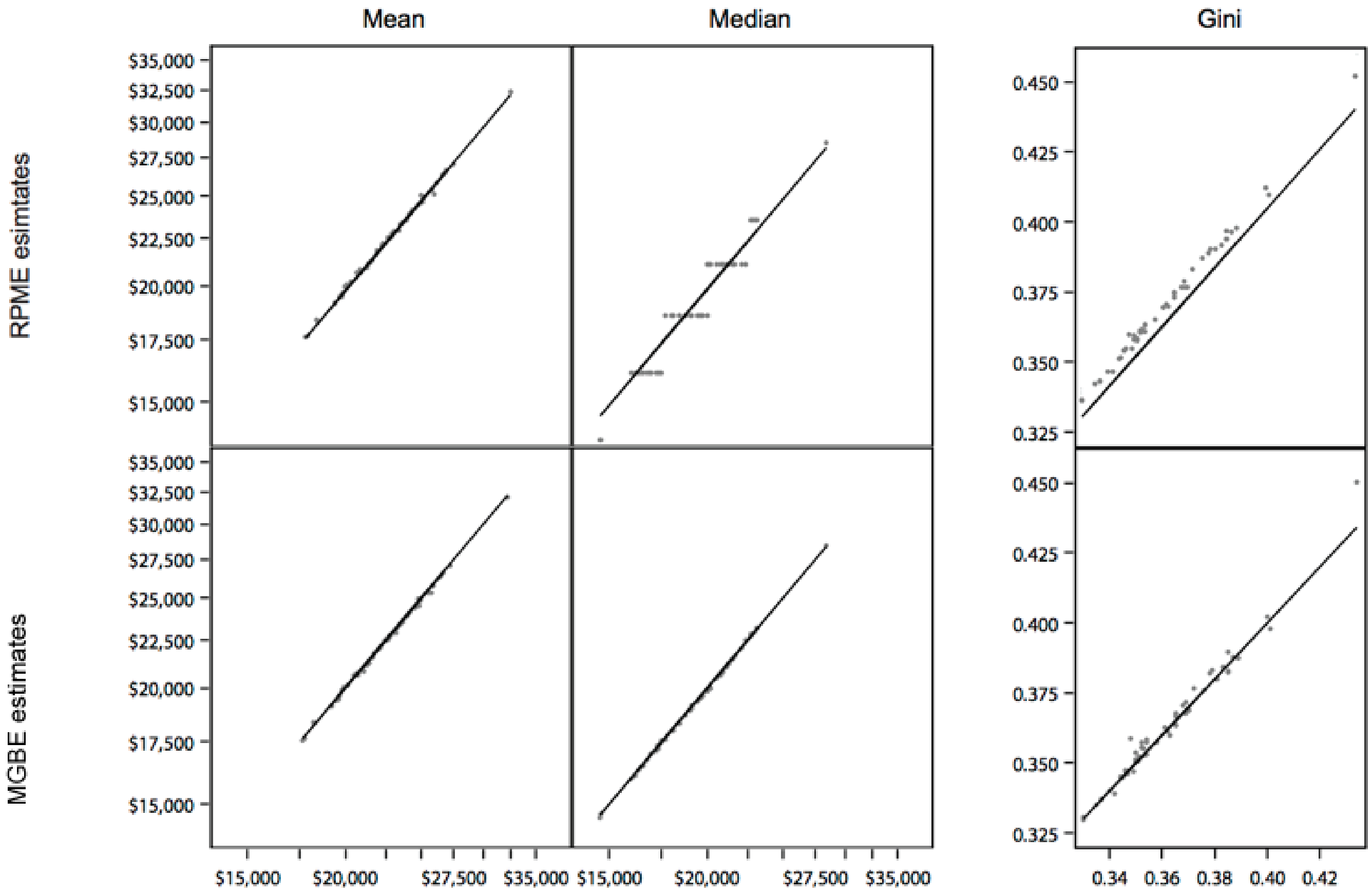

Figure 6. State estimates in 1980. If the estimates were perfectly accurate, all points would fall on the diagonal reference lines. When all the points are above the reference line, there is positive bias.

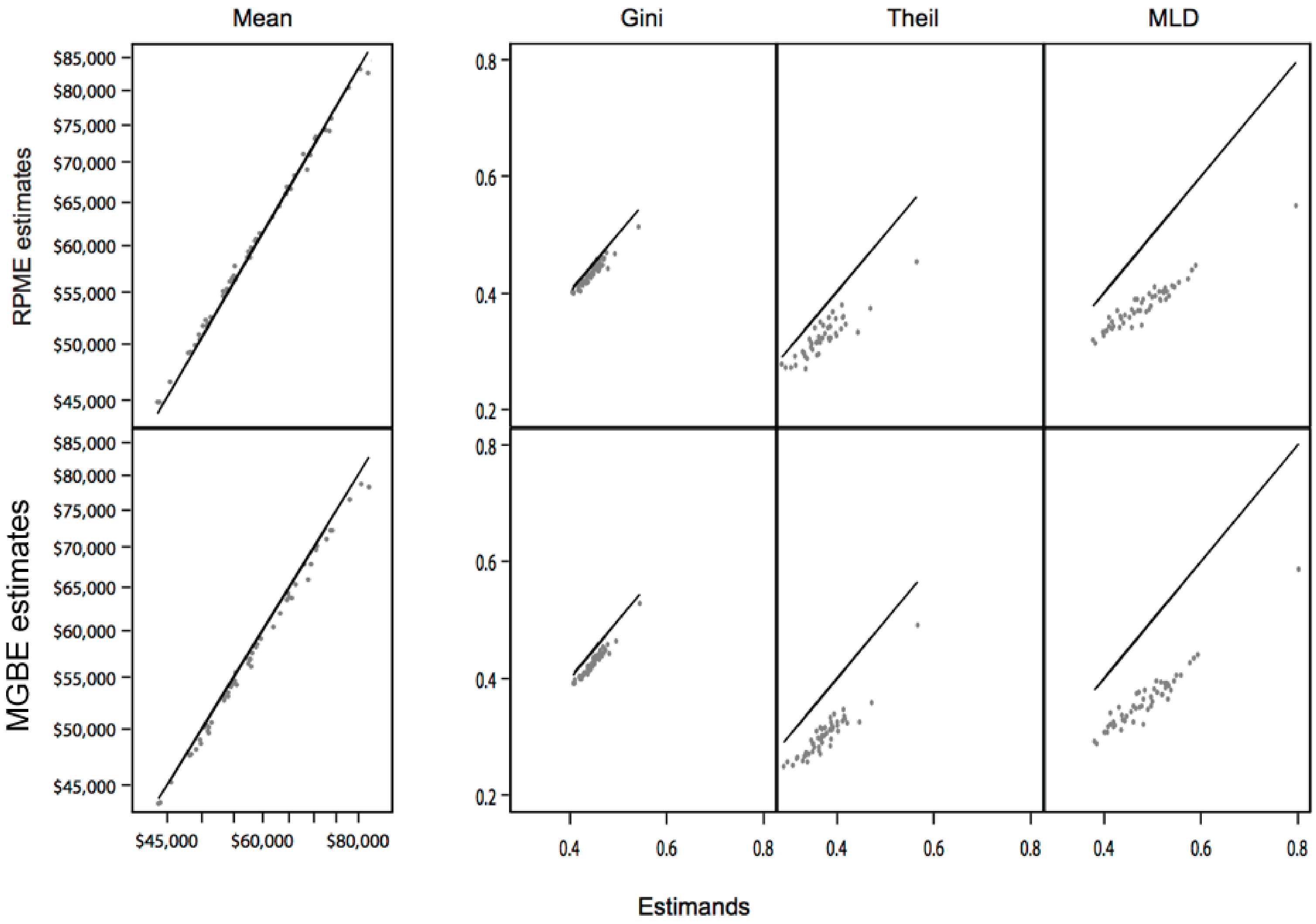

Figure 7. State estimates in 2005. If the estimates were perfectly accurate, all points would fall on the diagonal reference lines. When the points are below the reference line, there is negative bias.